\documentclass[12pt,a4paper,sort&compress]{elsarticle}
\usepackage[utf8x]{inputenc}
\usepackage{graphicx}
\usepackage{amsmath} 
\usepackage{amssymb} 

\usepackage[left=2cm,right=2cm,top=2cm,bottom=2cm]{geometry}
\usepackage[bookmarks=true,colorlinks=true,citecolor=blue,linkcolor=blue,
            urlcolor=magenta]{hyperref}

\usepackage{xcolor}

\usepackage{soul}
\sethlcolor{red}

\usepackage{multicol}

\journal{Progress in Nuclear Energy}

\begin{document}

\begin{frontmatter}

\title{Simulation of the traveling wave burning on epithermal neutrons on the year time scale}

\author[ONPU]{V.D.~Rusov\corref{cor1}}
\author[ONPU]{V.A.~Tarasov}
\author[CREST]{M.V.~Eingorn}
\author[ONPU]{S.A.~Chernezhenko}
\author[ONPU]{A.A.~Kakaev}
\author[ONPU]{V.P.~Smolyar}
\author[ONPU]{S.I.~Kosenko}
\author[ONPU]{T.N.~Zelentsova}

\cortext[cor1]{Corresponding author e-mail: siiis@te.net.ua}

\address[ONPU]{Department of Theoretical and Experimental Nuclear Physics, 
                           Odessa National Polytechnic University, 
                           Shevchenko av. 1, Odessa 65044, Ukraine}
\address[CREST]{CREST and NASA Research Centers, 
                            North Carolina Central University, 
                            Fayetteville st. 1801, Durham, North Carolina 27707, U.S.A.}

\begin{abstract}
We present the newly obtained results of two computer simulations of the epithermal neutron-nuclear burning in natural uranium. Each of them modeled the period of six months of the traveling wave reactor (TWR) operation -- for two different flux densities of an external neutron source.

The simulation results confirm the existence of a nuclear burning wave at longer time scales, and reveal the dependence of the wave burning modes on the parameters of an external neutron source.
\end{abstract}

\end{frontmatter}


\section{Introduction}

Following our previous work on numerical modeling of a wave burning in natural
uranium on epithermal neutrons~\cite{Rusov2015}, here we present another two
simulation results which confirm the possibility of the wave nuclear burning in
natural uranium in the epithermal region of neutron energies.

The first simulation reproduces the last example simulation
in~\cite{Rusov2015}, but for a much longer period of time. In this case the
neutron flux density was $10^{23}~neutron/cm^2 \cdot s$ and the simulated
reactor operation time was 150~days (as against 48~days in~\cite{Rusov2015}).

Another calculation was carried out for an external source with a significantly
lower flux density equal to $10^{15}~neutron/cm^2 \cdot s$ and for a simulation
time of 150~days.

\section{New simulation results of a wave neutron-nuclear burning in natural uranium on epithermal neutrons}

\begin{figure}[p!]
  \begin{minipage}[]{0.49\textwidth}
    \includegraphics[width=7cm]{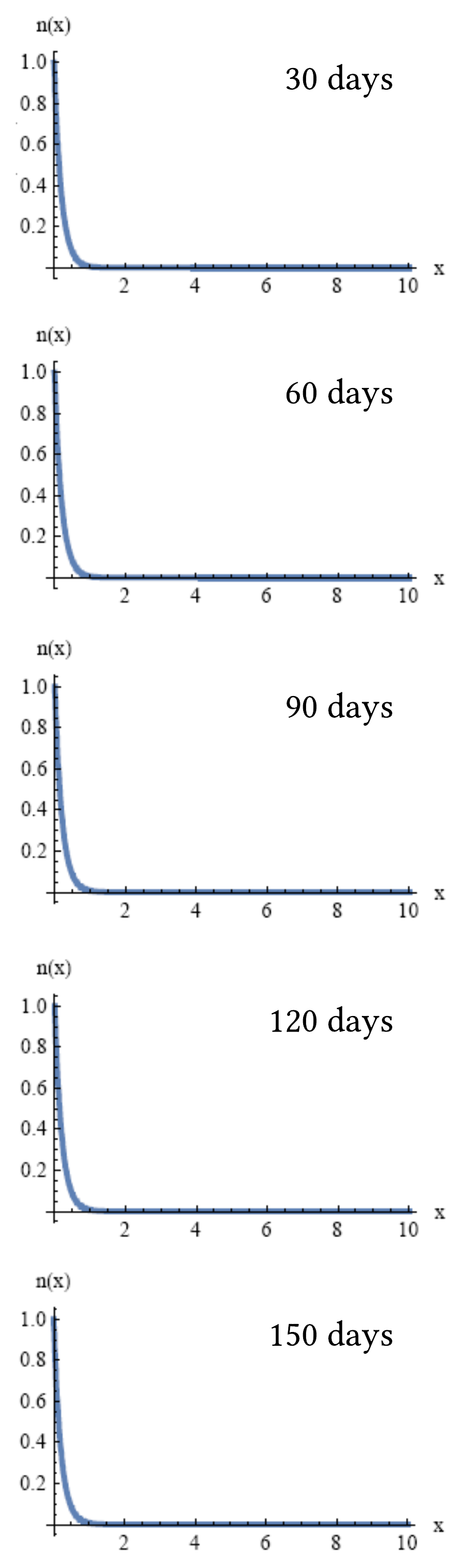}
    \caption{Kinetics of the neutrons density during the wave nuclear burning in
    natural uranium. The dimensionless neutron density is plotted.}
    \label{fig001}
  \end{minipage}
  \hfill
  \begin{minipage}[]{0.49\textwidth}
    \includegraphics[width=7cm]{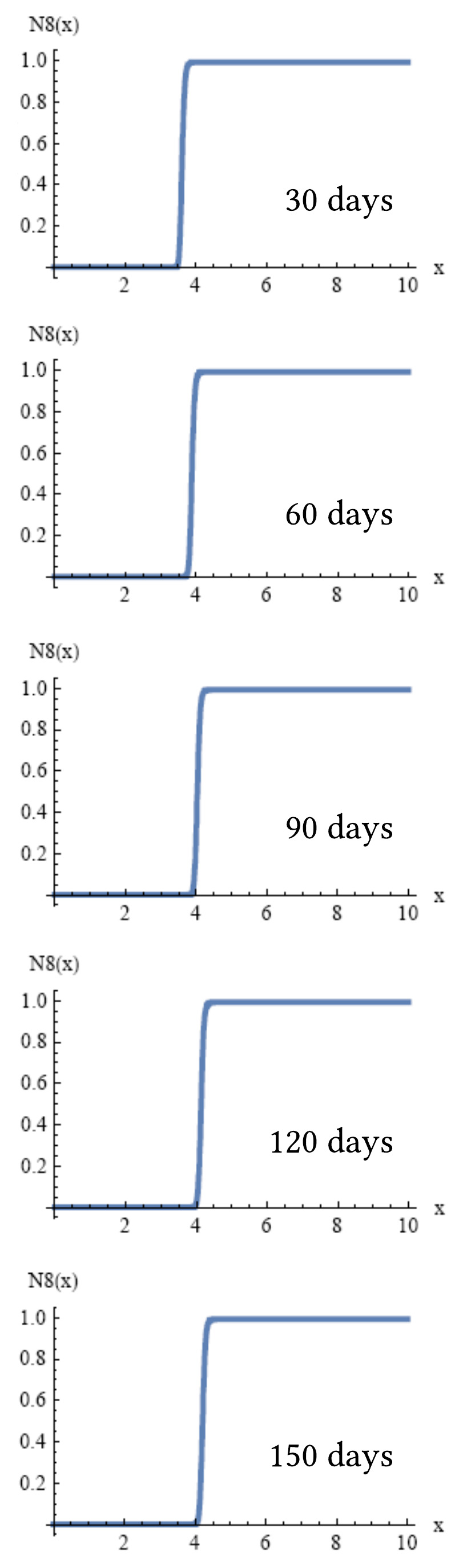}
    \caption{Kinetics of the $^{238}$U density during the wave nuclear burning in
    natural uranium. The dimensionless $^{238}$U density is plotted.}
    \label{fig002}
  \end{minipage}
\end{figure}

\begin{figure}[p!]
  \begin{minipage}[]{0.49\textwidth}
    \includegraphics[width=7cm]{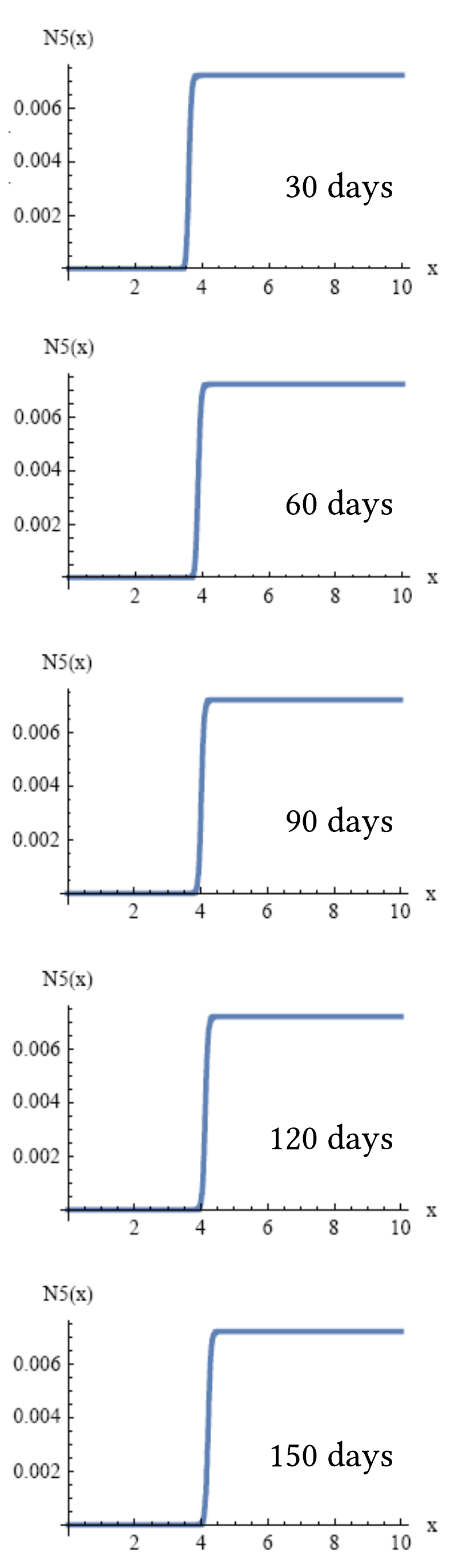}
    \caption{Kinetics of the $^{235}$U density during the wave nuclear burning in
    natural uranium. The dimensionless $^{235}$U density is plotted.}
    \label{fig003}
  \end{minipage}
  \hfill
  \begin{minipage}[]{0.49\textwidth}
    \includegraphics[width=7cm]{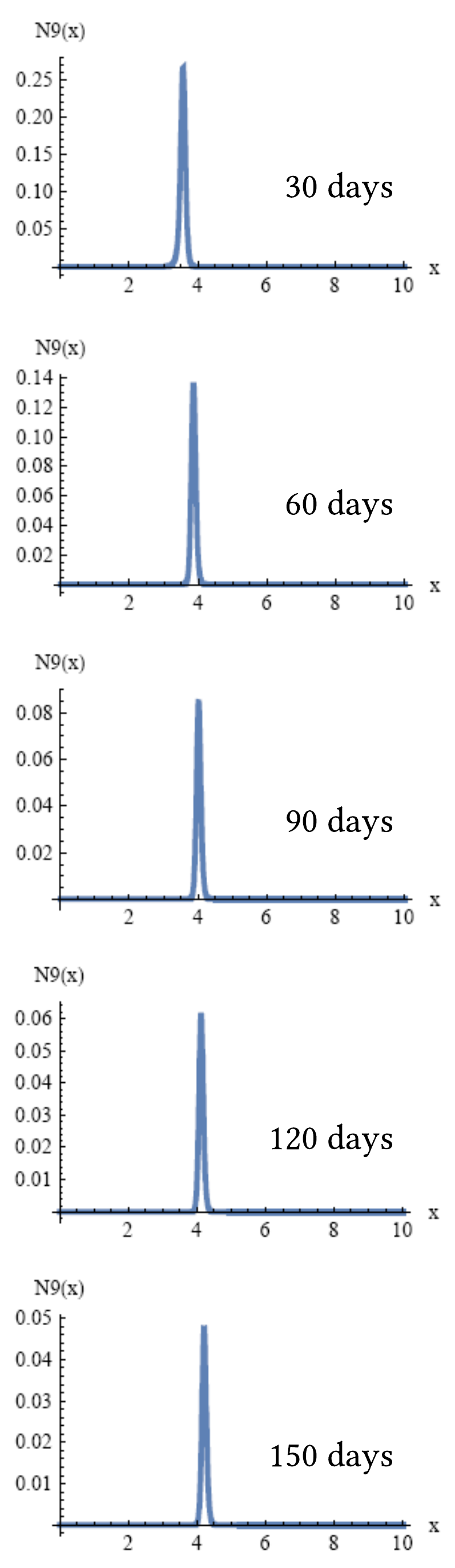}
    \caption{Kinetics of the $^{239}$U density during the wave nuclear burning in
    natural uranium. The dimensionless $^{239}$U density is plotted.}
    \label{fig004}
  \end{minipage}
\end{figure}

\begin{figure}[p!]
\begin{center}
\includegraphics[width=7cm]{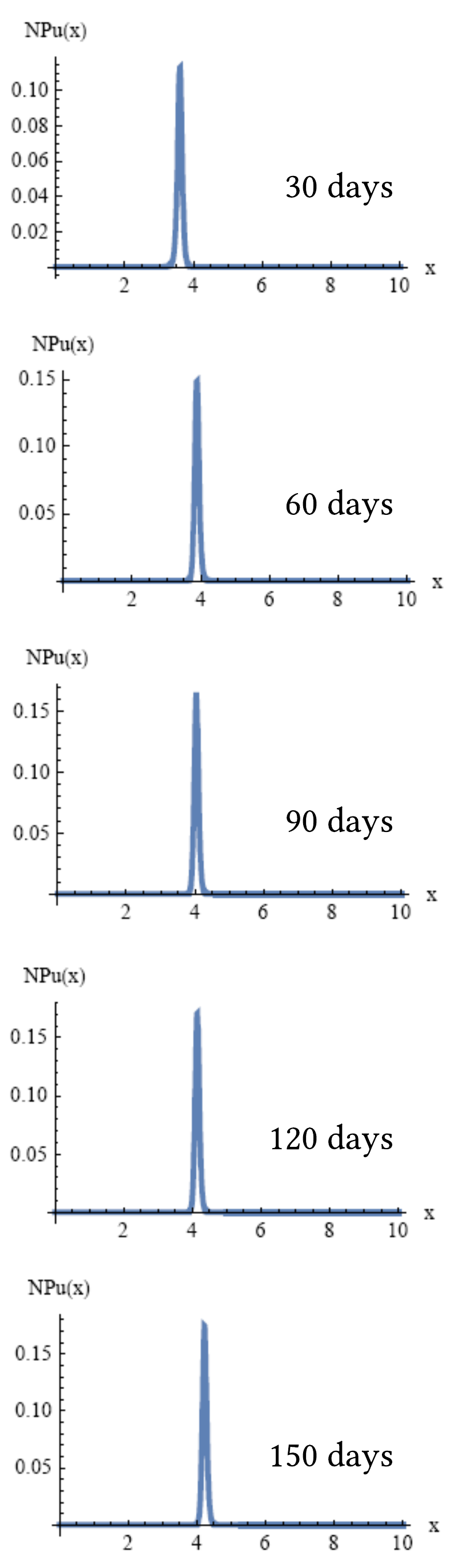}
\end{center}
\caption{Kinetics of the $^{239}$Pu density during the wave nuclear burning in
natural uranium. The dimensionless $^{239}$Pu density is plotted.}
\label{fig005}
\end{figure}

Figures~\ref{fig001}-\ref{fig005} show the simulation results for the neutron
flux density of an external source equal to $10^{23}~cm^{-2} \cdot s^{-1}$.
The kinetic system of 20 equations, the initial and boundary conditions
remained the same as in~\cite{Rusov2015}. The only difference here is the
simulated operation time.

In the present work, as in~\cite{Rusov2015}, the numerical solution of the
system of kinetic equations was carried out using the Mathematica~8 software
package. In order to optimize the process of numerical solution of the system
of equations, we switched to dimensionless variables 
$n(x,t) \rightarrow n^*(x,t)$ and $N(x,t) \rightarrow N^*(x,t)$, 
according to the following relations:
 
\begin{equation}
n(x,t) = \frac{\Phi_0}{V_n} n^* (x,t), ~ ~
N(x,t) = \frac{\rho_8 N_A}{\mu_8} N^* (x,t).
\label{eq001}
\end{equation}

In the first calculation, the following constants were used:

\begin{align}
& D = 2.0 \cdot 10^4 ~cm^2 / s; ~
V_n = 1.0 \cdot 10^6 ~cm / s; ~
\Phi_0 = 1.0 \cdot 10^{23} ~cm^{-2} s^{-1}; ~
\tau_\beta \sim 3.3 ~days;  \nonumber \\
& \nu ^{(Pu)} = 2.90; \nu^{(5)} = 2.41; ~ \nonumber \\
& \sigma_f ^{Pu} = 477.04 \cdot 10^{-24} ~cm^2; ~
\sigma_c ^{Pu} = 286.15 \cdot 10^{-24} ~cm^2; ~
\sigma _c^8 = 252.50 \cdot 10^{-24} ~cm^2; ~ \nonumber \\
& \sigma_f ^5 = 136.43 \cdot 10^{-24} ~cm^2; ~
\sigma_c ^5 = 57.61 \cdot 10^{-24} ~cm^2; ~
\sigma_c ^9 = 4.80 \cdot 10^{-24} ~cm^2; ~ \nonumber \\
& \sigma_c ^{eff (Pu)} = 10.10 \cdot 10^{-24} ~cm^2; ~
\sigma_c ^{i (Pu)} = 1.00 \cdot 10^{-24} ~cm^2 , i=1..6; ~ \nonumber \\
& \sigma_c ^{eff (5)} = 10.10 \cdot 10^{-24} ~cm^2; ~
\sigma_c ^{i (5)} = 1.00 \cdot 10^{-24} ~cm^2, i=1..6; \nonumber \\
\nonumber \\
& T_1 ^{(Pu)} = 54.28 ~s; ~
T_2 ^{(Pu)} = 23.04 ~s; ~
T_3 ^{(Pu)} = 5.60 ~s; ~ \nonumber \\
& T_4 ^{(Pu)} = 2.13 ~s; ~
T_5 ^{(Pu)} = 0.62 ~s; ~
T_6 ^{(Pu)} = 0.26 ~s; ~ \nonumber \\
& p_1 ^{(Pu)} = 0.072 \cdot 10^{-3}; ~
p_2 ^{(Pu)} = 0.626 \cdot 10^{-3}; ~
p_3 ^{(Pu)} = 0.444 \cdot 10^{-3}; ~ \nonumber \\
& p_4 ^{(Pu)} = 0.685 \cdot 10^{-3}; ~
p_5 ^{(Pu)} = 0.180 \cdot 10^{-3}; ~
p_6 ^{(Pu)} = 0.093 \cdot 10^{-3}; ~ \nonumber \\
& p^{(Pu)} = \sum \limits _{i=1} ^6 p_i ^{(Pu)} = 0.0021; ~
\sigma_c ^{eff} = 1.10 \cdot 10^{-24} ~cm^2; \nonumber \\
\nonumber \\
& T_1 ^{(5)} = 55.72 ~s; ~
T_2 ^{(5)} = 22.72 ~s; ~
T_3 ^{(5)} = 6.22 ~s; ~ \nonumber \\
& T_4 ^{(5)} = 2.30 ~s; ~
T_5 ^{(5)} = 0.61 ~s; ~
T_6 ^{(5)} = 0.23 ~s; ~ \nonumber \\
& p_1 ^{(5)} = 0.210 \cdot 10^{-3}; ~
p_2 ^{(5)} = 1.400 \cdot 10^{-3}; ~
p_3 ^{(5)} = 1.260 \cdot 10^{-3}; ~ \nonumber \\
& p_4 ^{(5)} = 2.520 \cdot 10^{-3}; ~
p_5 ^{(5)} = 0.740 \cdot 10^{-3}; ~
p_6 ^{(5)} = 0.27 \cdot 10^{-3}; ~ \nonumber \\
& p^{(5)} = \sum \limits _{i=1} ^6 p_i ^{(5)} = 0.0064 .
\label{eq002}
\end{align}

The length of the fissile medium, in which the wave of neutron-nuclear burning
propagates, is 1000~cm, the total simulation time is $t = 150~days$, the time
step is $\Delta t = 10~minutes$, and the spatial step is $\Delta x = 1~cm$.

As can be seen from the presented results, e.g. in Fig.~\ref{fig005}, after
90~days the amplitude of the $^{239}$Pu concentration reaches a stationary
maximum. This maximum of plutonium has shifted by 20~cm over 60 days, which
allows us to estimate the speed of steady-state wave burning, which is
approximately equal to $0.39 \cdot 10^{-5} ~cm/s$. It was not possible to
make such an estimation in~\cite{Rusov2015} because of the short simulation
time of 48 days.

\newpage
Below in Figures~\ref{fig006}-\ref{fig010} we show the simulation results for
the same kinetic system, the same basic calculation constants (see 
(\ref{eq002})), and the same initial and boundary conditions, except for the
value for the external neutron flux density, equal in this case to 
$10^{15} ~cm^{-2}s^{-1}$. The length of the fissile medium in which the wave of
neutron-nuclear burning propagates is 1000~cm, the total simulation time is
$t = 150~days$, the temporal step is $\Delta t = 5~minutes$, and the spatial
step is $\Delta x = 1~cm$.

\begin{figure}[p!]
  \begin{minipage}[]{0.49\textwidth}
    \includegraphics[width=7cm]{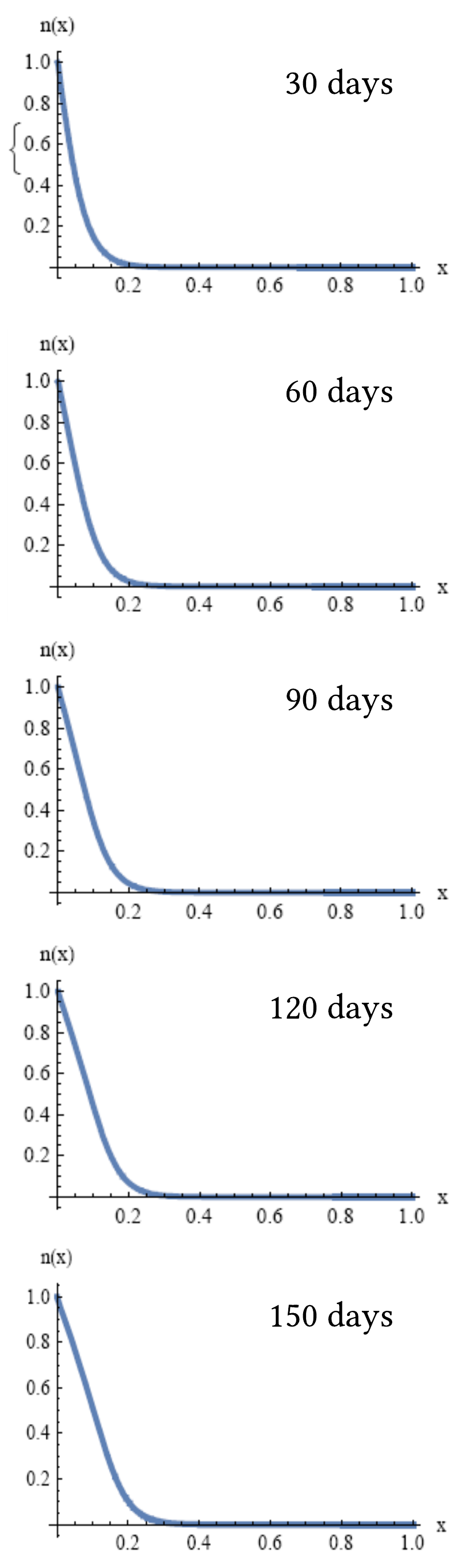}
    \caption{Kinetics of the neutrons density during the wave nuclear burning in
natural uranium for the external neutron flux of $10^{15} ~cm^{-2}s^{-1}$.
The dimensionless neutron density is plotted.}
    \label{fig006}
  \end{minipage}
  \hfill
  \begin{minipage}[]{0.49\textwidth}
    \includegraphics[width=7cm]{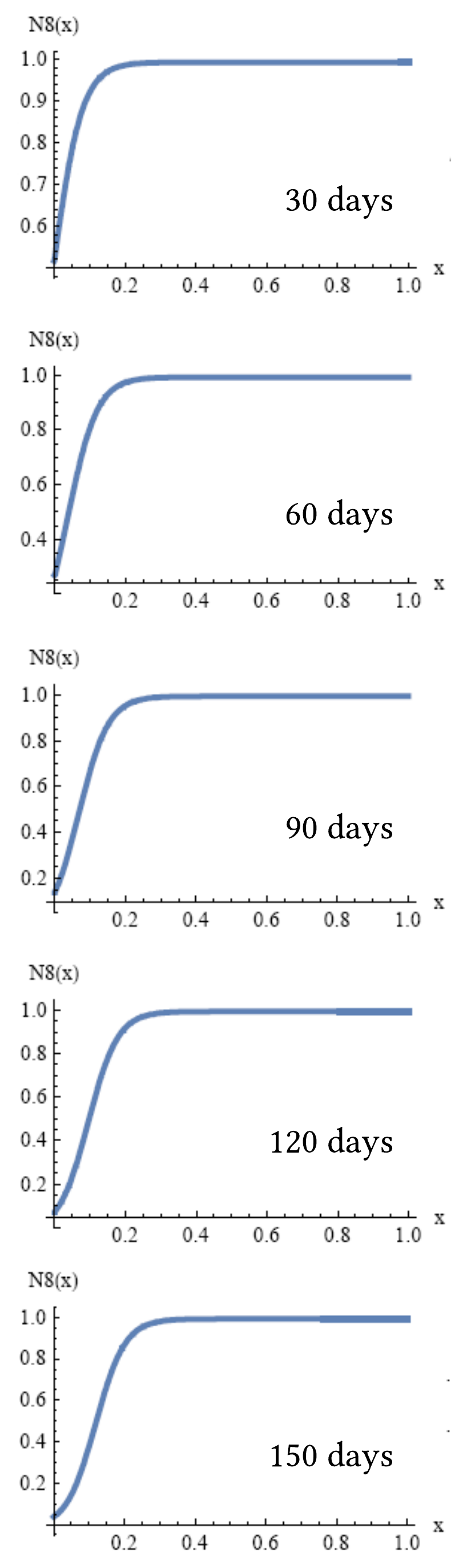}
    \caption{Kinetics of the $^{238}$U density during the wave nuclear burning in
natural uranium for the external neutron flux of $10^{15} ~cm^{-2}s^{-1}$.
The dimensionless $^{238}$U density is plotted.}
\label{fig007}
  \end{minipage}
\end{figure}

\begin{figure}[p!]
  \begin{minipage}[]{0.49\textwidth}
    \includegraphics[width=7cm]{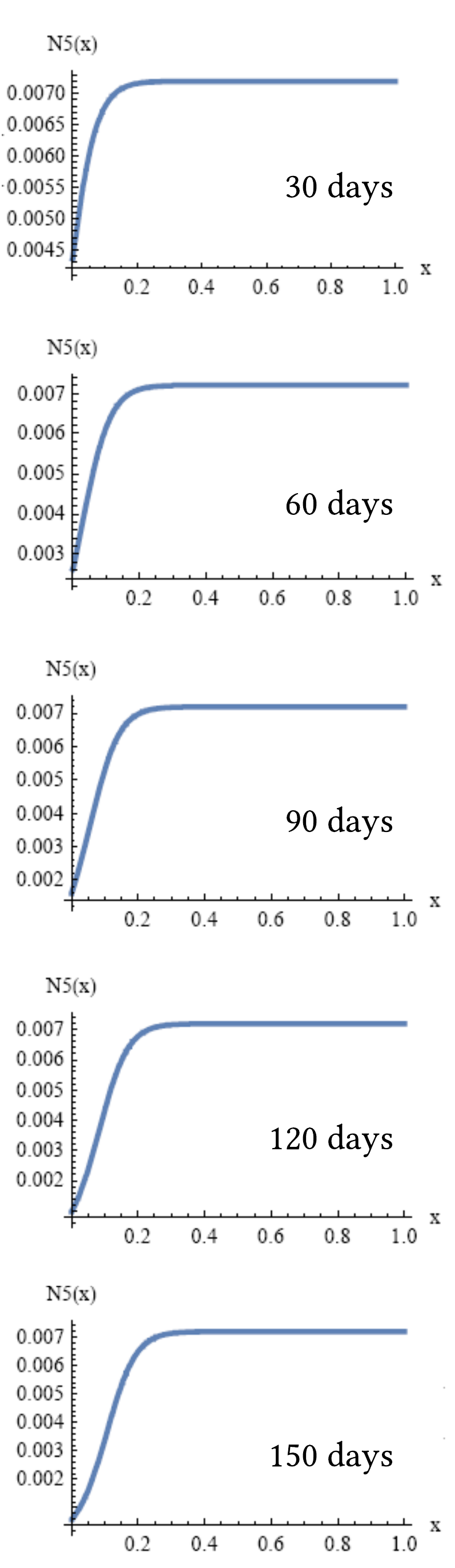}
    \caption{Kinetics of the $^{235}$U density during the wave nuclear burning in
natural uranium for the external neutron flux of $10^{15} ~cm^{-2}s^{-1}$.
The dimensionless $^{235}$U density is plotted.}
\label{fig008}
  \end{minipage}
  \hfill
  \begin{minipage}[]{0.49\textwidth}
    \includegraphics[width=7cm]{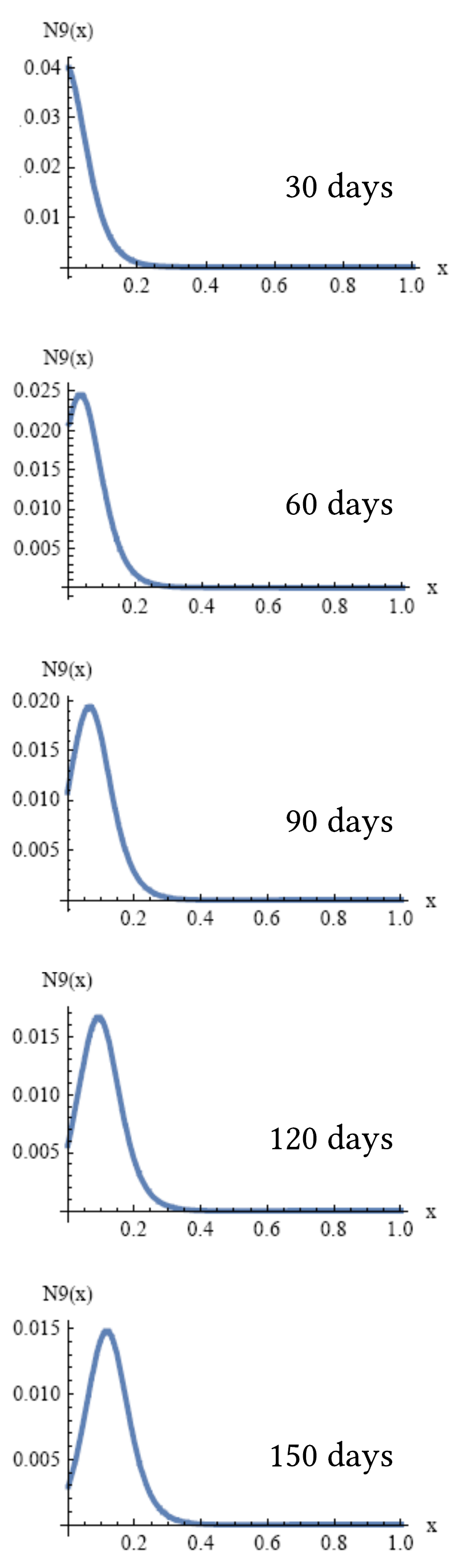}
    \caption{Kinetics of the $^{239}$U density during the wave nuclear burning in
natural uranium for the external neutron flux of $10^{15} ~cm^{-2}s^{-1}$.
The dimensionless $^{239}$U density is plotted.}
\label{fig009}
  \end{minipage}
\end{figure}

\begin{figure}[p!]
\begin{center}
\includegraphics[width=7cm]{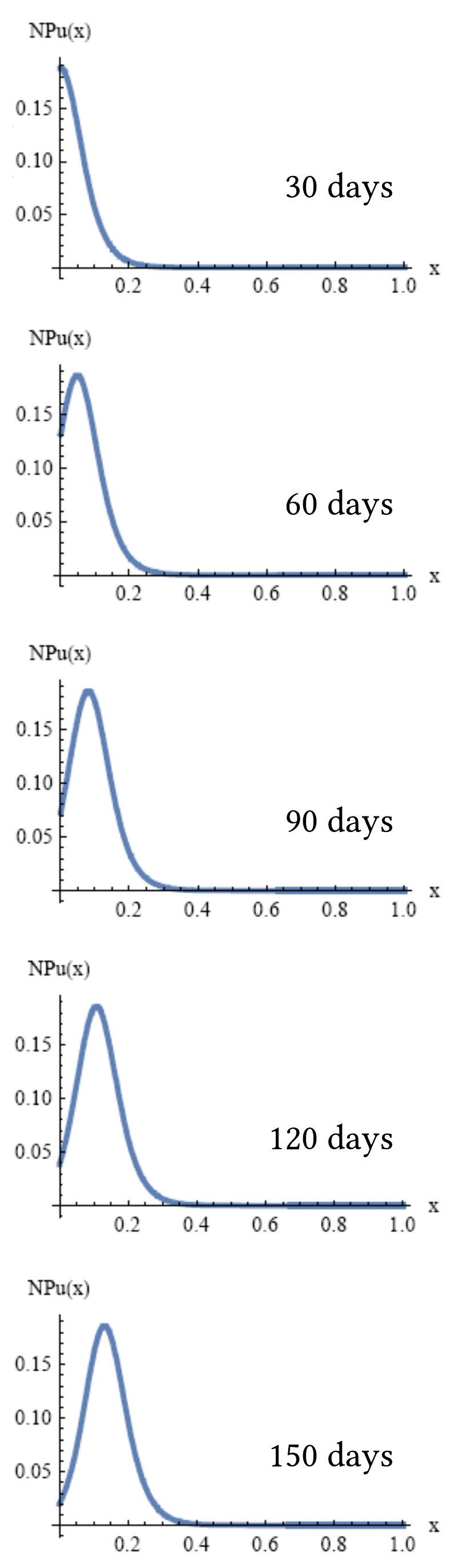}
\end{center}
\caption{Kinetics of the $^{239}$Pu density during the wave nuclear burning in
natural uranium for the external neutron flux of $10^{15} ~cm^{-2}s^{-1}$.
The dimensionless $^{239}$Pu density is plotted.}
\label{fig010}
\end{figure}

From the presented results, e.g. Fig.~\ref{fig010}, it can be seen that after
60 days the amplitude of the $^{239}$Pu concentration reaches a steady maximum,
which shifts by 60~cm during the simulated period of 90 days. This allows us to
estimate the speed of steady-state wave burning, which is approximately equal
to
\begin{equation}
u_{calc} \approx 60~cm / (90 \cdot 24 \cdot 3600~s) \approx 0.77 \cdot 10^{-5} ~cm/s .
\label{eq006}
\end{equation}

A comparative analysis of the results of these two numerical experiments
shows that with a decrease in the neutron flux density of the external source
from $10^{23} ~cm^{-2}s^{-1}$ to $10^{15} ~cm^{-2}s^{-1}$, all other constants
unchanged (the only difference is the time step of 5min in the latter case),
the wave burning speed approximately doubled, and the wave half-width also
increased significantly (evaluation was carried out using $^{239}$Pu) from
10~cm to 100~cm.

An explanation of the change in the parameters of the burning regime, such as
the phase velocity and the width of the wave burning zone, when only the flux
density of the external source is changed, can be given if we recall the
quantum-mechanical analogy for the diffusion equation for neutrons. The
requirement to fulfill the Bohr-Somerfeld quantization rule for solutions of
this equation put forward by L.P.~Feoktistov in~\cite{Feoktistov1989}, studied
in~\cite{Ershov2003}, and later developed by V.D.~Rusov~\textit{et~al.} into
the quantum-statistical Wigner distribution for phase velocities of the wave
burning e.g. in~\cite{Rusov2011,RusovSTNI2015}.

Indeed, in~\cite{Rusov2011,RusovSTNI2015} it allowed to write the critical
condition for steady-state wave burning in the form:

\begin{equation}
I = \int \sqrt{\dfrac{n^{Pu239}}{n_{crit}^{Pu239}} - 1} dx = \frac{\pi}{2} , 
\label{eq003}
\end{equation}

\noindent
where the integral is taken over the supercritical region 
$n^{Pu239} > n_{crit}^{Pu239}$.

As the simulation results show, even when only the flux density of the external
source changes, within our simplified kinetic system, the width of the burning
region changes, and according to (\ref{eq003}), the ratio between $n^{Pu239}$
and $n_{crit}^{Pu239}$ must also change, since the integral should remain
constant.

Hence we can make a conclusion about the change in the equilibrium-stationary
and critical concentrations of $^{239}$Pu which are the parameters in the
Wigner quantum-statistical distribution for phase velocities of the wave
burning~\cite{Rusov2011,RusovSTNI2015}, and this causes the change in the wave
burning speed.

Note that the presented simulation results clearly demonstrate the influence of
the parameters of an external neutron source on the parameters of the burning
regime even for the simplified model under consideration. In a real process of
nuclear burning, not only the flux density of an external neutron source, but
also its energy spectrum will have a significant effect on the burning regime.

According to the theory of a soliton-like neutron wave of slow nuclear burning,
developed on the basis of the quantum chaos
theory~\cite{Rusov2011,RusovSTNI2015}, the neutron-nuclear burning speeds must
satisfy the Wigner quantum-statistical distribution. The phase velocity $u$ of
a soliton-like neutron wave of nuclear burning is determined by the following
approximate equality:

\begin{equation}
\Lambda (a_*) = \frac{u \tau_\beta}{2L} \cong \left( \frac{8}{3 \sqrt{\pi}} \right) ^6 a_* ^4 \exp {(-\frac{64}{9 \pi} a_* ^2)}, ~ ~
a_* ^2 = \frac{\pi ^2}{4} \cdot \frac{N_{crit}^{Pu}}{N_{eq}^{Pu} - N_{crit}^{Pu}},
\label{eq004}
\end{equation}

\noindent
where $\Lambda (a_*)$ is a dimensionless invariant, depending on the parameter
$a_*$; $N_{eq}^{Pu}$ and $N_{crit}^{Pu}$ are the equilibrium and critical
concentrations of $^{239}$Pu, $L$ is the mean free path of neutrons, 
$\tau_\beta$ is the delay time, associated with the production of the active
(fissile) isotope and equal to the effective period of the $\beta$-decay of
compound nuclei in the Feoktistov uranium-plutonium cycle.

To check the correspondence between the Wigner distribution (\ref{eq004}) and
the phase velocity of the slow neutron-nuclear burning of natural uranium in
the epithermal region of neutron energies, which is obtained by numerical
simulation, let us make the corresponding estimates of the parameter $a_*^2$
and invariant $\Lambda (a_*)$.

For this we can use the data of numerical simulation at an external source
flux density of $10^{15} ~cm^{-2}s^{-1}$, shown in Fig.~\ref{fig010} for 
$^{239}$Pu.

From fig.~\ref{fig010}, one can find 
$N_{eq}^{Pu} \approx 0.19 \times 4.8 \cdot 10^{22} ~cm^{-3}$ (maximum on the
$^{239}$Pu concentration curve) and 
$N_{crit}^{Pu} \approx 0.1 \times 4.8 \cdot 10^{22} ~cm^{-3}$ (inflection point
on the $^{239}$Pu concentration curve).

Then, according to (\ref{eq004}), for the parameter $a_*$ and
invariant $\Lambda (a_*)$ we obtain the following estimates:

\begin{equation}
a_* = \sqrt{\frac{\pi^2}{4} \frac{0.1 \cdot 10^{22}~cm^{-3}}
{0.19 \cdot 10^{22}}~cm^{-3} - 0.1\cdot 10^{22}~cm^{-3}} = 
\sqrt{\frac{\pi^2}{4} \frac{0.1}{0.09}} \approx 1.66
~~~and~~~
\Lambda (a_*) \approx 0.3 .
\label{eq005}
\end{equation}

\begin{figure}[tb!]
\begin{center}
\includegraphics[width=15cm]{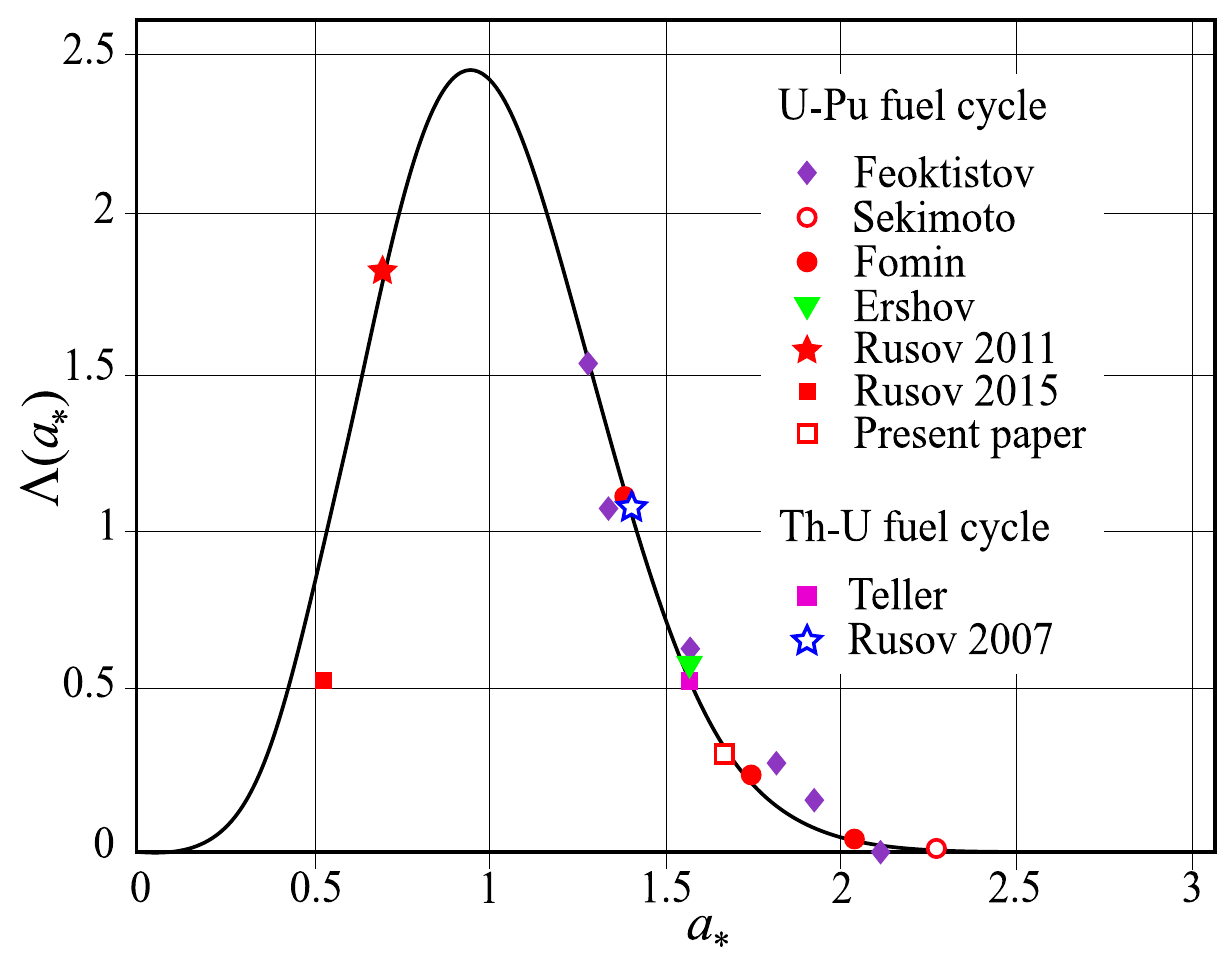}
\end{center}
\caption{Theoretical (solid line) and calculated (points) dependence for the
phase velocity of neutron-nuclear burning $\Lambda (a_*) = u \tau_\beta / 2L$
on the parameter $a_*$. Data combined 
from~\cite{Teller1995,Teller1996,RusovJGR2007,Sekimoto2006,Fomin2005,Fomin2007,
Rusov2011,RusovCHAOS2011,RusovSTNI2015,Rusov2015,Pavlovich2007}, and
supplemented by the estimate obtained in the present paper for slow wave
burning of natural uranium in the epithermal region of neutron energies
(1.0 - 7.0~eV).}
\label{fig011}
\end{figure}

These estimates for $a_*$ and $\Lambda (a_*)$ are shown in Fig.~\ref{fig011}.

In order to estimate the phase velocity of the wave burning from these
parameters, we first have to estimate the mean free path for epithermal
neutrons in the studied medium.

The mean free path for neutrons of the indicated epithermal region of
neutron energies is:

\begin{equation}
L = \frac{1}{\Sigma_a} = \frac{1}{\bar{\sigma}_c ^9 N_8 (t=0)} \approx 
\frac{1}{4.68 \cdot 10^{-24} ~cm^2 \cdot 0.48 \cdot 10^{23} ~cm^{-3}} \approx 
4.45 ~cm.
\label{eq007}
\end{equation}

Then from expression (\ref{eq004}) we can also obtain an estimate for the
phase velocity of burning:

\begin{equation}
u = \frac{2 \cdot \Lambda \cdot L}{\tau_\beta} \approx 
\frac{2 \cdot 0.3 \cdot 4.45~cm}{2.85 \cdot 10^5 ~s} \approx
0.94 \cdot 10^{-5} ~cm/s .
\label{eq008}
\end{equation}

A comparison of (\ref{eq006}) and (\ref{eq008}) demonstrates a rather good
agreement.

It is also possible to estimate the heat power produced by the reactor with
the active zone of natural uranium, provided the wave burning mode is realized
in it.

Indeed, given that the derivative of the $^{239}$Pu concentration with respect
to time and the derivative with respect to the coordinate are related by the
following equation:

\begin{equation}
\frac{dN^{Pu239}}{dt} = u \frac{dN^{Pu239}}{dx} ,
\label{eq010}
\end{equation}

\noindent
where $u$ is the wave burning speed for the case of external source with
neutron flux density of $10^{15} ~cm^{-2}s^{-1}$, according to 
Fig.~\ref{fig010}, it is possible to calculate the specific heat power
$P_{spec}$:

\begin{align}
P_{spec} & \approx 210.3 ~MeV \cdot \frac{dN^{Pu239}}{dt} = 
210.3 ~MeV \cdot u \frac{dN^{Pu239}}{dx} \approx \nonumber \\
& \approx 210.3 ~MeV \cdot 0.77 \times 10^{-5} \times 0.48 \cdot 10^{22} 
\frac{1}{cm^3 \cdot s} \approx 
77.73 \cdot 10^{17} \frac{MeV}{cm^3 \cdot s} \approx \nonumber \\
& \approx 77.73 \cdot 10^{23} \times 1.6 \cdot 10^{-19} \frac{J}{cm^3} \approx
1.24 \frac{MW}{cm^3} .
\label{eq011}
\end{align}

For an example active zone in a form of a cylinder with 50cm diameter, the
estimate of the burning zone volume is:

\begin{equation}
V_{burning} \approx \pi \cdot r^2 \cdot L \approx 
3.14 \cdot (25 ~cm)^2 \cdot 4.45 ~cm \approx 8730 ~cm^3
\label{eq012}
\end{equation}

And the resulting heat power is:

\begin{equation}
P = P_{spec} \cdot V_{burning} \approx 1.24 \frac{MW}{cm^3} 8730 ~cm^3 
\approx 10.8 ~GW.
\label{eq013}
\end{equation}

It should be noted that the simulated kinetics of neutron density presented in
Figs.~\ref{fig001} and~\ref{fig006}, does not demonstrate the neutron wave, in
contrast to the previously published results for the $^{238}U$ wave burning in
fast neutron range (neutron energy 
$\sim 1 ~MeV$)~\cite{Rusov2011,RusovSTNI2015}.

We believe that this is related to the use of a permanent external source of
neutrons, as in~\cite{Rusov2015}, with a rather high flux density, which hides
the neutron wave on its background.

\section{Conclusions}

We presented new results of two computer simulations of the epithermal
neutron-nuclear burning in natural uranium for two different densities of the
external neutron flux. Each simulation represents about half a year of reactor
operation.

The obtained results clearly demonstrate the influence of the external neutron
source parameters on the burning mode even for the simplified model under
consideration.

In a real process of nuclear burning, not only the flux density of an external
neutron source, but also its energy spectrum will have a significant effect on
the burning regime.

Based on the simulation results, we estimate the wave burning speed and the
corresponding heat power production of the epithermal wave nuclear reactor with
a homogeneous cylindrical core with a diameter of 50~cm made of natural uranium
and a moderator. The resulting power of such reactor is about 10~GW.

\section*{Acknowledgements}

M.V. Eingorn acknowledges support by NSF CREST award HRD-0833184 and NASA grant 
NNX09AV07A and thanks Prof. Branislav Vlahovic (NCCU) for the given 
computational resources without which the numerical calculation would be 
impossible on the same level of accuracy.

\bibliography{Tarasov-EpithermalUPu}

\end{document}